\documentclass[conference,a4paper]{IEEEtran}

\usepackage{graphicx,epic,eepic,epsfig,amsmath,latexsym,amssymb,verbatim,color,revsymb}
\usepackage{theorem}
\usepackage{cite}

\usepackage[usenames,dvipsnames]{pstricks}

\newtheorem{definition}{Definition}
\newtheorem{proposition}[definition]{Proposition}
\newtheorem{lemma}[definition]{Lemma}

\newtheorem{theorem}[definition]{Theorem}

\def\squareforqed{\hbox{\rlap{$\sqcap$}$\sqcup$}}
\def\qed{\ifmmode\squareforqed\else{\unskip\nobreak\hfil
\penalty50\hskip1em\null\nobreak\hfil\squareforqed
\parfillskip=0pt\finalhyphendemerits=0\endgraf}\fi}
\def\endenv{\ifmmode\;\else{\unskip\nobreak\hfil
\penalty50\hskip1em\null\nobreak\hfil\;
\parfillskip=0pt\finalhyphendemerits=0\endgraf}\fi}
\newcommand{\altqed}{\hfill{\small $\blacksquare$}}
\mathchardef\ordinarycolon\mathcode`\:
\mathcode`\:=\string"8000
\def\vcentcolon{\mathrel{\mathop\ordinarycolon}}
\begingroup \catcode`\:=\active
  \lowercase{\endgroup
  \let :\vcentcolon
  }

\newcommand{\nc}{\newcommand}
\nc{\rnc}{\renewcommand}
\nc{\beq}{\begin{equation}}
\nc{\eeq}{{\end{equation}}}
\nc{\beqa}{\begin{eqnarray}}
\nc{\eeqa}{\end{eqnarray}}
\nc{\beas}{\begin{eqnarray*}}
\nc{\eeas}{\end{eqnarray*}}
\nc{\lbar}[1]{\overline{#1}}
\nc{\bra}[1]{\langle#1|}
\nc{\ket}[1]{|#1\rangle}
\nc{\ketbra}[2]{|#1\rangle\!\langle#2|}
\nc{\braket}[2]{\langle#1|#2\rangle}
\nc{\proj}[1]{| #1\rangle\!\langle #1 |}
\nc{\avg}[1]{\langle#1\rangle}
\nc{\Rank}{\operatorname{Rank}}
\nc{\smfrac}[2]{\mbox{$\frac{#1}{#2}$}}
\nc{\tr}{\operatorname{Tr}}
\nc{\ox}{\otimes}
\nc{\dg}{\dagger}
\nc{\dn}{\downarrow}
\nc{\cA}{{\cal A}}
\nc{\cB}{{\cal B}}
\nc{\cC}{{\cal C}}
\nc{\cD}{{\cal D}}
\nc{\cE}{{\cal E}}
\nc{\cF}{{\cal F}}
\nc{\cG}{{\cal G}}
\nc{\cH}{{\cal H}}
\nc{\cI}{{\cal I}}
\nc{\cJ}{{\cal J}}
\nc{\cK}{{\cal K}}
\nc{\cL}{{\cal L}}
\nc{\cM}{{\cal M}}
\nc{\cN}{{\cal N}}
\nc{\cO}{{\cal O}}
\nc{\cP}{{\cal P}}
\nc{\cR}{{\cal R}}
\nc{\cS}{{\cal S}}
\nc{\cT}{{\cal T}}
\nc{\cX}{{\cal X}}
\nc{\cZ}{{\cal Z}}
\nc{\csupp}{{\operatorname{csupp}}}
\nc{\qsupp}{{\operatorname{qsupp}}}
\nc{\var}{{\operatorname{var}}}
\nc{\rar}{\rightarrow}
\nc{\lrar}{\longrightarrow}
\nc{\polylog}{{\operatorname{polylog}}}
\nc{\1}{{\openone}}
\nc{\wt}{{\operatorname{wt}}}
\nc{\av}[1]{{\left\langle {#1} \right\rangle}}

\nc{\RR}{{{\mathbb R}}}
\nc{\CC}{{{\mathbb C}}}
\nc{\FF}{{{\mathbb F}}}
\nc{\NN}{{{\mathbb N}}}
\nc{\ZZ}{{{\mathbb Z}}}
\nc{\PP}{{{\mathbb P}}}
\nc{\QQ}{{{\mathbb Q}}}
\nc{\UU}{{{\mathbb U}}}
\nc{\EE}{{{\mathbb E}}}
\nc{\id}{{\operatorname{id}}}

\nc{\CHSH}{{\operatorname{CHSH}}}

\nc{\be}{\begin{equation}}
\nc{\ee}{{\end{equation}}}
\nc{\bea}{\begin{eqnarray}}
\nc{\eea}{\end{eqnarray}}
\nc{\<}{\langle}
\rnc{\>}{\rangle}
\nc{\Hom}[2]{\mbox{Hom}(\CC^{#1},\CC^{#2})}
\nc{\rU}{\mbox{U}}

\nc{\ob}[1]{#1}

\nc{\SEP}{{\text{SEP}}}
\nc{\sep}{{\text{sep}}}
\nc{\LOCC}{{\text{LOCC}}}
\nc{\PPT}{{\text{PPT}}}
\nc{\EXT}{{\text{EXT}}}
\nc{\Sym}{{\operatorname{Sym}}}
\nc{\SWAP}{{\operatorname{SWAP}}}

\newcommand{\bes}{\begin{equation*}}
\newcommand{\ees}{\end{equation*}}

\newcommand{\semistrong}{{pretty strong}}
\newcommand{\Semistrong}{{Pretty strong}}

\begin{document}

\title{``\Semistrong'' converse for the private capacity\protect\\ of degraded quantum wiretap channels}

\author{\IEEEauthorblockN{Andreas Winter}
%\thanks{Dated 24 April 2016.}%
\IEEEauthorblockA{ICREA and Departament de F\'{\i}sica, 
                  Grup d'Informaci\'{o} Qu\`{a}ntica\\
                  Universitat Aut\`{o}noma de Barcelona,
                  ES-08193 Bellaterra (Barcelona), Spain.\\
%                  Email: {\tt andreas.winter@uab.cat}
}}

\maketitle

\begin{abstract}
In the vein of the recent ``\semistrong{}'' converse for the quantum
and private
capacity of degradable quantum channels [Morgan/Winter, IEEE Trans. Inf.
Theory 60(1):317-333, 2014], we use the same techniques, in particular
the calculus of min-entropies, 
to show a \semistrong{} converse for the private capacity of degraded 
classical-quantum-quantum (cqq-)wiretap channels, 
which generalize Wyner's model of the degraded 
classical wiretap channel.

While the result is not completely tight, leaving some gap between 
the region of error and privacy parameters for which the converse bound
holds, and a larger no-go region, it represents a further step
towards an understanding of strong converses of wiretap channels
[cf. Hayashi/Tyagi/Watanabe, arXiv:1410.0443 for the classical case].
\end{abstract}


\maketitle

\section{Introduction}
One of the most outstanding successes of Shannon theory~\cite{Shannon48}
is Shannon's information theoretic treatment of cryptography~\cite{Shannon49}, 
and the further development at the hands of Wyner, who introduced the 
wiretap channel model~\cite{Wyner:wiretap}. While the achievability 
part for the wiretap channel is a well-understood combination of
channel coding and privacy amplification techniques, the converse,
even the weak converse, of the generalized wiretap channel required
a new idea in Csisz\'{a}r and K\"orner's contribution~\cite{CsiszarKoerner:wiretap}.

Characteristically, for multi-user scenarios strong converses are
hard to come by and not known in many instances. The presence of an
adversary in the wiretap setting, albeit a passive one, makes the
wiretap capacity a multi-user problem, and until recently only
weak converses were known for Wyner's original 
problem~\cite{Wyner:wiretap,CsiszarKoerner:wiretap}. 
The same was is true for the ``static'' versions of 
distillation of shared secret key between Alice and Bob from a prior 
three-way correlation with Eve~\cite{Maurer,AhlswedeCsiszar}, where
Tyagi and Narayan~\cite{TyagiNarayan}, 
Tyagi and Watanabe~\cite{Tyagi}, and 
Watanabe and Hayashi~\cite{WatanabeHayashi}
made progress only recently.
Most recently, Hayashi, Tyagi and Watanabe~\cite{HayashiTyagiWatanabe}
(see also their~\cite{HTW-0})
have given an elegant, very insightful analysis of strong converse
rates for general classical wiretap channels, yielding the complete
strong converse in the degraded case.

Here, we extend their results somewhat to the quantum case, looking
at wiretap channels with classical input but quantum outputs,
so-called cqq-wiretap channels. Instead of the elegant hypothesis testing
method developed in~\cite{HayashiTyagiWatanabe}, we use a rather
more blunt tool, the min-entropy calculus~\cite{Renner:PhD,TomamichelThesis}.
Hence, while we can treat channels not amenable to the method 
of~\cite{HayashiTyagiWatanabe}, we do not reach a complete understanding
of the full tradeoff between decoding error and privacy.

\section{Cqq-wiretap channel and strong converse}
The model we consider is that of a discrete memoryless cqq-wiretap channel:
\begin{align*}
  W:\cX &\longrightarrow \cS(B\ox E) \\
      x &\longmapsto     \rho_x^{BE},
\end{align*}
with a finite set $\cX$ and finite dimensional Hilbert spaces $B$ and $E$,
of legal user and eavesdropper, respectively.
Furthermore, we shall assume most of the time that the channel is
\emph{degraded}, meaning that there exists a quantum channel (cptp 
map) $\cD:\cL(B) \rightarrow \cL(E)$ such that $\rho_x^E = \cD(\rho_x^B)$
for all $x\in\cX$. Introducing a Stinespring dilation of $\cD$ by an
isometry $V:B \hookrightarrow E'\ox F$, we have $\rho_x^{E'} = \tr_F V\rho_x^B V^\dagger$.

The objective of wiretap coding is for Alice to encode messages in
such a way that Bob can decode with small error probability, and
that Eve cannot distinguish messages except with small probability.
To quantify errors, we use the \emph{purified distance}
\[
  P(\rho,\sigma) = \sqrt{1-F(\rho,\sigma)^2},
\]
with the fidelity $F(\rho,\sigma) = \| \sqrt{\rho}\sqrt{\sigma} \|_1$
between quantum states~\cite{Uhlmann:fid,Jozsa:fid}, see~\cite{TomamichelThesis}.

An \emph{$n$-block code of transmission error $\epsilon$ and privacy error $\delta$}
for $W$ consists of a stochastic map $E:[M]\longrightarrow \cX^n$ and
a POVM $D=(D_u)_{u=1}^M$ on $B^n$, such that for
\[\begin{split}
  \rho^{U\widehat{U}E^n} 
       = \frac{1}{M}\sum_{u,\hat{u},x^n} E(x^n|u)& \proj{u}^U \ox \proj{\hat{u}}^{\widehat{U}} \\
                                                 &\phantom{=}
                                                  \ox \tr_{B^n}\bigl[\rho_{x^n}^{B^nE^n}(D_{\hat{u}}\ox\1)\bigr],
\end{split}\]
the following hold:
\begin{align}
  P(\rho^{U\widehat{U}},\Delta^{U\widehat{U}})     &\leq \epsilon, \\
  P(\rho^{UE^n},\Delta^U\ox\widetilde{\rho}^{E^n}) &\leq \delta.
\end{align}
Here, $\Delta^{U\widehat{U}} = \frac1M \sum_u \proj{u}^U \ox \proj{u}^{\widehat{U}}$,
so that $\Delta^U$ is the maximally mixed state, and 
$\widetilde{\rho}^{E^n}$ is a suitable state on $E^n$.

The largest number $M$ of messages under these conditions is
denoted $M(n,\epsilon,\delta)$. 
Then, the \emph{private capacity} is defined as
the largest asymptotically achievable rate such that transmission
error and privacy error vanish in the limit, i.e.
\[
  P(W) := \inf_{\epsilon,\delta>0} \liminf_{n\rightarrow\infty} \frac1n \log M(n,\epsilon,\delta).
\]

\begin{theorem}[{Devetak~\cite{Devetak}; Cai/Winter/Yeung~\cite{CaiWinterYeung}}]
  \label{thm:P}
  Let $W$ be
  a cqq-wiretap channel. Then its private capacity is given by
  $P(W) = \sup_n \frac1n P^{(1)}(W^{\ox n})$, where
  \[
    P^{(1)}(W) = \max I(U:B) - I(U:E).
  \]
  Here, the maximum is over joint distributions $P_{UX}$ of the
  channel input $X$ and an auxiliary variable $U$, and the mutual informations 
  are with respect to the state
  \[
    \rho^{UXBE} = \sum_{u,x} P_{UX}(u,x) \proj{u}^U \ox \proj{x}^X \ox \rho_x^{BE}.
  \]
  For degraded channels, it is given by the single-letter formula
  \[
    P(W) = P^{(1)}(W) = \max I(X:F|E'),
  \]
  where the maximum is over distributions $P_X$ of the channel input,
  and the conditional mutual information is with respect to the state
  \[
    \rho^{XE'FE} = \sum_x P_X(x) \proj{x}^X (V\ox\1) \rho_x^{BE} (V\ox\1)^\dagger.
  \]
  In other words, w.l.o.g.~one may assume $U=X$, and the regularization is not 
  necessary~\cite[Appendix~A]{Enigma}.
  \altqed
\end{theorem}

\medskip
For completeness, we recall here the definition of the quantum information
quantities: For a state $\rho$ on a quantum system $X$, the entropy is
$S(X) = S(\rho^X) = -\tr\rho\log\rho$, the mutual information for
a bipartite state $\rho^{XY}$ is $I(X:Y) = S(X)+S(Y)-S(XY)$, and
the conditional mutual information for a tripartite state $\rho^{XYZ}$ 
is $I(X:Y|Z) = S(XZ)+S(YZ)-S(Z)-S(XYZ)$.

\medskip
It seems to be unknown whether in the cqq-wiretap channel setting
the regularization above is necessary, but it is quite clear
that the single-letterization in the classical case, by Csiszar 
and K\"orner~\cite{CsiszarKoerner:wiretap}, does not work, 
due to the use of chain rules, we would get information quantities
conditioned on quantum registers. Furthermore, the results of
Smith, Renes and Smolin~\cite{SmithRenesSmolin08} suggest
that $P^{(1)}$ does not give the private capacity.
On the other hand, in the general quantum channel case~\cite{CaiWinterYeung}
it is well-known that the regularization is necessary: For an 
isometry, such that Eve's channel is the complementary channel to Bob's, 
there are instances where $P^{(1)}$ is strictly smaller than 
$P$~\cite{SmithRenesSmolin08,LWZG09,SmithSmolin:P,ElkoussStrelchuk}.
Also if the eavesdropper's channel is a degraded version (even trivially)
of the authorized channel, and the latter is quantum, $P^{(1)}$
can be strictly smaller than $P$, as observed in~\cite{Enigma}.

Here we show the following \semistrong{} converse:
\begin{theorem}
  \label{thm:P-prettystrong}
  Let $W : \cX \rightarrow \cS(B\ox E)$ be a degraded cqq-wiretap channel
  and $\epsilon,\delta \geq 0$ such that $\epsilon + 2\delta < 1$,
  then
  \[
    \log M(n,\epsilon,\delta) \leq n P(W) + O\bigl(\sqrt{n\log n}\bigr),
  \]
  where the implicit constant only depends on $1-\epsilon-2\delta$.
%  where $C=...$ and 
%  $\eta = \frac16\left(1-\epsilon-2\delta\right)$.
  In particular, under the above assumptions,
  \[
    \lim_{n\rightarrow\infty} \frac{1}{n}\log M(n,\epsilon,\delta) = P(W).
  \]
\end{theorem}

Its proof relies on the calculus of min- and max-entropies,
of which we will briefly review the necessary definitions and properties; 
cf.~\cite{TomamichelThesis} for more details.

\begin{definition}[Min- and max-entropy]
For $\rho^{AB} \in \cS_{\leq}(AB)$,
the min-entropy of $A$ conditioned on $B$ is defined as
\[
  H_{\min}(A|B)_{\rho} := \max_{\sigma_B \in \mathcal{S}(B)} 
                          \max \{\lambda \in \mathbb{R} : \rho^{AB} \leq 2^{-\lambda} \1 \otimes \sigma^B \}.
\]
With a purification $\ket{\psi}^{ABC}$ of $\rho$, we define
\[
  H_{\max}(A|B)_{\rho} := - H_{\min}(A|C)_{\psi^{AC}},
\]
with the reduced state $\psi^{AC} = \tr_B \psi$.
\end{definition}

\begin{definition}[Smooth min- and max-entropy]
\label{def:smoothed}
Let $\epsilon \geq 0$ and $\rho_{AB} \in \mathcal{S}({AB})$. The 
\emph{$\epsilon$-smooth min-entropy of $A$ conditioned on $B$} is defined as
\[
  H^{\epsilon}_{\min}(A|B)_{\rho} 
     := \max_{{\rho'} \approx_{\epsilon} \rho} H_{\min}(A|B)_{\rho'},
\]
where ${\rho'} \approx_{\epsilon} \rho$ means $P(\rho',\rho) \leq \epsilon$ for
$\rho'\in\cS_{\leq}(AB)$.

Similarly,
\[\begin{split}
  H_{\max}^{\epsilon}(A|B)_{\psi} 
     &:= \min_{{\rho'} \approx_{\epsilon} \rho} H_{\max}(A|B)_{\rho'}\\
     &=  - H_{\min}^{\epsilon}(A|C)_{\psi},
\end{split}\]
with a purification $\psi \in \mathcal{S}({ABC})$ of $\rho$.

All min- and max-entropies, smoothed or not, are invariant under local
unitaries and local isometries.
\end{definition}

The following two lemmas show that min- and max-entropies
have properties close to those of the von Neumann entropy.
\begin{lemma}[Data processing]
\label{lemma:mono}
For $\rho\in\cS(ABC)$ and $\epsilon \geq 0$,
\begin{align*}
  H_{\min}^\epsilon(A|BC) &\leq H_{\min}^\epsilon(A|B), \\
  \phantom{=======:}
  H_{\max}^\epsilon(A|BC) &\leq H_{\max}^\epsilon(A|B).
  \phantom{========}{\small\blacksquare}
\end{align*}
%\altqed
\end{lemma}

\begin{lemma}[Chain rules~\cite{DBWR10,VDTR12}]
\label{lemma:chain-rules}
Let $\epsilon,\delta\geq 0$, $\eta >0$. Then, with respect to the same 
state $\rho \in \mathcal{S}({ABC})$,
\begin{equation}\begin{split}
  \label{eq:chain:max-le-max+max}
  H_{\max}^{\epsilon + 2\delta + \eta}\!(AB|C)
     &\!\leq\! H_{\max}^{\delta}\!(B|C) \!+\! H_{\max}^{\epsilon}\!(A|BC) 
%     &\phantom{=============}
      + \log\!\frac{2}{\eta^2},
\end{split}\end{equation}
%and
\begin{equation}\begin{split}
  \label{eq:chain:max-ge-min+max}
  \phantom{=}
  H_{\max}^{\epsilon}\!(AB|C)
     &\!\geq\! H_{\min}^{\delta}\!(B|C) \!+\! H_{\max}^{\epsilon+2\delta+2\eta}\!(A|BC) 
%     &\phantom{=============}
      - 3\log\!\frac{2}{\eta^2}.
      \phantom{:}{\small\blacksquare}
\end{split}\end{equation}
%\altqed
\end{lemma}

\begin{IEEEproof}[Proof of Theorem~\ref{thm:P-prettystrong}]
We follow closely the initial steps of the analysis in~\cite[Thm.~14]{MorganWinter}.
Consider an $n$-block code with $M$ messages, and transmission and privacy error 
$\epsilon$ and is $\delta$: message $u$ (chosen 
uniformly) is encoded by a distribution $E(x^n|u)$ and sent through the 
channel, giving rise to a ccqq-state between message $U$, 
input $X^n$, output $B^n$ and environment $E^n$:
\[
  \rho^{UX^nB^nE^n} = \frac{1}{M}\sum_{u,x^n} E(x^n|u) \proj{u}^U \ox \proj{x^n}^{X^n} \ox \rho_{x^n}^{B^nE^n}.
\]
The ``trivial'' converse shows that
\[
  \log M \leq H_{\min}^\delta(U|E^n) - H_{\max}^\epsilon(U|B^n),
\]
cf.~Renes and Renner~\cite{RenesRenner}. Namely, according 
to the definition of privacy given above, the reduced state $\rho^{UE^n}$
is within purified distance $\delta$ of a product state of the form 
$\frac{1}{M}\sum_u \proj{u}^U \ox \widetilde{\rho}^{E^n}$, hence 
$H_{\min}^\delta(U|E^n) \geq \log M$. Likewise, there exists a decoding cptp
map $\cD:\cL(B^n) \rightarrow \widehat{U}$ such that $(\id\ox\cD)\rho^{UB^n}$ 
is within $\epsilon$ purified distance from the perfectly correlated state
$\frac{1}{M}\sum_u \proj{u}^U \ox \proj{u}^{\widehat{U}}$, hence
$H_{\max}^\epsilon(U|B^n) \leq 0$. 

We apply the Stinespring dilation of the degrading map to $\rho$, yielding
\[\begin{split}
  &\omega^{UX^n{E'}^nF^nE^n} \\
  &\phantom{:}
   = \frac{1}{M}\sum_{u,x^n} E(x^n|u) 
                    \proj{u}^U \ox \proj{x^n}^{X^n} \ox V^{\ox n} \rho_{x^n}^{B^nE^n} {V^{\dagger}}^{\ox n}.
\end{split}\]
With respect to this state, we now have (cf.~Eq.~(18) of \cite{MorganWinter}),
\begin{equation}\begin{split}
  \label{eq:privacy-bound}
  \log M &\leq H_{\min}^\delta(U|E^n) - H_{\max}^\epsilon(U|{E'}^nF^n) \\
         &=    H_{\min}^\delta(U|{E'}^n) - H_{\max}^\epsilon(U|{E'}^nF^n) \\
         &\leq H_{\max}^\eta(F^n|{E'}^n) - H_{\max}^{\epsilon+2\delta+5\eta}(F^n|{E'}^n U) 
                                                                   + 4\log\frac{2}{\eta^2} \\
         &\leq H_{\max}^\eta(F^n|{E'}^n) - H_{\max}^{\lambda}(F^n|{E'}^n X^n) 
                                                                   + 4\log\frac{2}{\eta^2},
\end{split}\end{equation}
with $\lambda := \epsilon+2\delta+5\eta$. 
Here we have used the degradability property of the channel in the second line,
in the third line twice the chain rule for min-/max-entropies
[Lemma~\ref{lemma:chain-rules}, Eqs.~(\ref{eq:chain:max-le-max+max}) 
and (\ref{eq:chain:max-ge-min+max})], 
and in the last line data processing (Lemma~\ref{lemma:mono})
for the max-entropy.
Indeed,
\begin{align*}
  H_{\max}^{\epsilon+3\eta}(AB|C) &\leq H_{\max}^\eta(A|C) + H_{\max}^\epsilon(B|AC) 
                                                                                + \log\frac{2}{\eta^2} \\
    \parallel \phantom{===}       &                                                                    \\
  H_{\max}^{\kappa}(AB|C)         &\geq H_{\min}^{\delta}(B|C) + H_{\max}^{\kappa+2\delta+2\eta}(A|BC) 
                                                                               - 3\log\frac{2}{\eta^2},
\end{align*}
which we apply with $F^n \equiv A$, $U\equiv B$ and ${E'}^n \equiv C$,
and with $\kappa = \epsilon+3\eta$.

Now, assume for simplicity that the distribution of $X^n$, i.e.~the
density $\omega^{X^n}$, is supported on a single type class
\[
  \tau(P_0) = \bigl\{ x^n : \forall x\ F(x|x^n) = P_0(x) \bigr\},
\]
where $F(x|x^n)$ is the relative requency of the letter $x$ 
occurring in $x^n$.
With this assumption we show how to bound the max-entropy terms 
on the r.h.s.~of Eq.~(\ref{eq:privacy-bound}).
Namely, all $x^n$ having non-zero probability are permutations of
a fiducial $x_0^n$. Hence, on the one hand, we have
\[\begin{split}
  H_{\max}^{\lambda}(F^n|{E'}^n X^n)_\omega 
       &\geq H_{\max}^{\widehat{\lambda}}(F^n|{E'}^n)_{\rho_{x_0^n}} \\
       &\geq n\sum_x P_0(x) S(F|E')_{\rho_x} - O\bigl(\sqrt{n}\bigr),
\end{split}\]
by Lemma~\ref{lemma:concavity} (quasi-concavity of max-entropy) below,
with $\widehat{\lambda} = \lambda\sqrt{2-\lambda^2} < 1$,
and a simple extension of Proposition~\ref{prop:AEP} 
(asymptotic equipartition property) to a product of blocks
of i.i.d.~states $\rho_x^{\ox nP_0(x)}$.
On the other hand, going to the $S_n$-symmetrized state
\[
  \Omega^{X^n{E'}^nF^nE^n}
     = \frac{1}{|\tau(P_0)|}\!\!\sum_{x^n \in \tau(P_0)}\!\!\!
                    \proj{x^n}^{X^n} \ox V^{\ox n} \rho_{x^n}^{B^nE^n} {V^{\dagger}}^{\ox n}\!\!,
\]
we have, once more invoking Lemma~\ref{lemma:concavity},
\[\begin{split}
  H_{\max}^\eta(F^n|{E'}^n)_\omega 
       &\leq H_{\max}^{\eta/\sqrt{2}}(F^n|{E'}^n)_\Omega \\
       &\leq H_{\min}^{1-\frac18 \eta^2}(F^n|{E'}^n)_\Omega,
\end{split}\]
where in the second line we have used Lemma~\ref{lemma:max-min-inequality}.
Now, the uniform distribution $\Upsilon_{P_0}$ on the type class $\tau(P_0)$ 
has the property
$\Upsilon_{P_0} \leq (n+1)^{|\cX|} P_0^{\ox n}$, because
it is on $P_0$ that the probability weight of type classes of
$P_0^{\ox n}$ peaks, and so
\[
  \Omega \leq (n+1)^{|\cX|} \bigl( \Theta^{XE'FE} \bigr)^{\ox n},
\]
with
\[
  \Theta^{XE'FE} = \sum_x P_0(x) \proj{x}^X \ox V \rho_x^{BE} V^\dagger.
\]
Thus, using the same reasoning as in the proof of~\cite[Thm.~2]{MorganWinter},
we get
\[\begin{split}
  H_{\max}^\eta(F^n|{E'}^n)_\omega 
       &\leq H_{\min}^{1-\frac{1}{16} \eta^2 (n+1)^{-|\cX|}}(F^n|{E'}^n)_{\Theta^{\ox n}} \\
       &\leq H_{\max}^{\frac{1}{32} \eta^2 (n+1)^{-|\cX|}}(F^n|{E'}^n)_{\Theta^{\ox n}} \\
       &\leq n S(F|E')_\Theta + O\bigl(\sqrt{n\log n}\bigr),
\end{split}\]
where we have once more invoked the AEP, Proposition~\ref{prop:AEP}.

Inserting these upper and lower bounds into Eq.~(\ref{eq:privacy-bound}),
we find
\[
  \log M \leq n I(X:F|E') + O\bigl(\sqrt{n\log n}\bigr).
\]

\medskip
Now we face the case of general encodings, and reduce it to the
above form of constant type. Introduce another register
$T$ holding the type $t(x^n)$ of $x^n$, of dimension $|T| \leq (n+1)^{|\cX|}$,
so that we have an extended joint state
\[\begin{split}
  \rho^{UX^nB^nE^nT} 
     = \frac{1}{M}\sum_{u,x^n} &E(x^n|u) \proj{u}^U \ox \proj{x^n}^{X^n} \\
                               &\phantom{==}
                                \ox \rho_{x^n}^{B^nE^n} \ox \proj{t(x^n)}^T.
\end{split}\]
Imagine that $T$ is handed to the eavesdropper; this clearly doesn't 
increase Bob's decoding error, but it can affect the privacy of the code.
The idea is, however, that since $\log|T| \leq O(\log n)$, we can
rectify this by hashing out $O(\log n)$ of the message, and the 
remainder will be almost as private as the original code:
Indeed, let $\epsilon' = \epsilon+2\vartheta$ and $\delta=\delta+2\vartheta$,
such that still $\epsilon'+2\delta' < 1$, however.
Partition $[M]$ randomly into $N = \left\lfloor \frac{M}{L} \right\rfloor$
sets $\cL_{u'}$ of equal size 
$L=\text{poly}\bigl(\log n,\vartheta^{-1}\bigr)$,
up to a rest of size smaller than $L$, so that we can label the
elements of $\bigcup_{u'} \cL_{u'}$ by pairs $(u',v) \in [N]\times[L]$.
Compute $U'$ as a function of $U$ (except for an event of
probability $\leq \frac1N$), so that we obtain a joint state
\[\begin{split}
  \widetilde{\rho}^{U'X^nB^nE^nT} 
     = \frac{1}{N}\sum_{u',v,x^n} &\frac1L E(x^n|u',v) \proj{u'}^{U'} \ox \proj{x^n}^{X^n} \\
                                  &\phantom{====}
                                   \ox \rho_{x^n}^{B^nE^n} \ox \proj{t(x^n)}^T             \\
     = \frac{1}{N}\sum_{u',x^n} &\widetilde{E}(x^n|u') \proj{u'}^{U'} \ox \proj{x^n}^{X^n} \\
                                &\phantom{=}
                                 \ox \rho_{x^n}^{B^nE^n} \ox \proj{t(x^n)^T},
\end{split}\]
where $\widetilde{E}(x^n|u') = \frac1L \sum_{v} E(x^n|u',v)$.
This is a new code: By the properties of random hashing,
with high probability, Bob can apply the same decoding as in the 
original code to obtain an error $\leq \epsilon+\vartheta$,
and the privacy error for the combined register $E^n T$ is 
$\leq \delta+\vartheta$. Furthermore, the privacy error of the register
$T$ alone is $\leq \vartheta$. This has the important consequence that
we can modify the encoding $\widetilde{E}(x^n|u')$ to a slightly different
one $E'(x^n|u') = Q(t(x^n)) E'(x^n|u',t(x^n))$, with a universal
distribution $Q$ over the types, such that 
\[\begin{split}
  {\rho'}^{U'X^nB^nE^nT} 
     = \frac{1}{N}\sum_{u',x^n} &E'(x^n|u') \proj{u'}^{U'} \ox \proj{x^n}^{X^n} \\
                                &\phantom{===}
                                 \ox \rho_{x^n}^{B^nE^n} \ox \proj{t(x^n)}      \\
     = \frac{1}{N}\sum_{P_0 \text{ type}} \sum_{u',x^n\in\tau(P_0)} 
                                &E'(x^n|u',P_0) \proj{u'}^{U'} \ox \proj{x^n}^{X^n} \\
                                &\phantom{====}
                                 \ox \rho_{x^n}^{B^nE^n} \ox Q(P_0) \proj{P_0}^T
\end{split}\]
fulfills the decoding and eavesdropper constraints with transmission error $\epsilon'$
and privacy error $\delta'$, and has a perfectly independent type-register $T$.

Consider now the codes obtained by using $E'(\cdot|\cdot,P_0)$ for a fixed 
$P_0$ (but always the same decoder for Bob). These have transmission errors
$\epsilon(P_0)$ and privacy errors $\delta(P_0)$. By the direct sum over
types $P_0$ -- with probability $Q(P_0)$ --, and the concavity of $\sqrt{1-x^2}$,
one can see that
\begin{equation*}
  \epsilon' \geq \sum_{P_0 \text{ type}} Q(P_0) \epsilon(P_0), \quad
  \delta'   \geq \sum_{P_0 \text{ type}} Q(P_0) \delta(P_0), 
\end{equation*}
and so
\[
  \sum_{P_0 \text{ type}} Q(P_0) [\epsilon(P_0) + 2\delta(P_0)] \leq \epsilon'+2\delta' < 1,
\]
and so there must exist a type $P_0$ such that the encoding $E'(\cdot|\cdot,P_0)$
has $\epsilon(P_0) + 2\delta(P_0) \leq \epsilon'+2\delta' < 1$. 
But this code has only $O(\log n)$ less information in the message, and
has the property that the encoder maps only into
the type class $\tau(P_0)$, hence can use the previous bound:
\[
  \log M \leq \log N + O(\log n) \leq n I(X:F|E') + O\bigl(\sqrt{n\log n}\bigr),
\]
concluding the proof.
\end{IEEEproof}

%\medskip
%Here follow the auxiliary results required in the proof.

\begin{lemma}[Lemma~10 in~\cite{MorganWinter}]
\label{lemma:concavity}
Let $\rho \in \cS(AB)$ be a state and consider the state family
$\rho_i^{AB} = (U_i\ox V_i)\rho(U_i\ox V_i)^\dagger$, with unitaries $U_i$ on $A$
and $V_i$ on $B$, and probabilities $p_i$; define
$\overline{\rho} := \sum_i p_i \rho_i$. Then,
with $\widehat{\epsilon} = \epsilon\sqrt{2-\epsilon^2} \leq \epsilon\sqrt{2}$,
\[
  \phantom{=======:}
  H_{\max}^{\epsilon}(A|B)_{\overline{\rho}} \geq H_{\max}^{\widehat{\epsilon}}(A|B)_{\rho}.
  \phantom{=======:}
  {\small\blacksquare}
\]
\end{lemma}

\begin{proposition}[Min- and max-entropy AEP \cite{Renner:PhD,TomamichelThesis}]
\label{prop:AEP}
Let $\rho \in \mathcal{S}(\mathcal{H}_{AB})$ and $0 < \epsilon < 1$. 
Then, 
\[\begin{split}
  \lim_{n \to \infty} \frac{1}{n} H^{\epsilon}_{\min}(A^n | B^n)_{\rho^{\ox n}} 
                    &= S(A|B)_{\rho}                                           \\
                    &= \lim_{n \to \infty} \frac{1}{n} H^{\epsilon}_{\max}(A^n | B^n)_{\rho^{\ox n}}.
\end{split}\]
%where $S(A|B) = S(AB)-S(B)$ is the conditional von Neumann entropy.

More precisely, for a purification $\ket{\psi} \in ABC$ of $\rho$, denote 
$\mu_X := \log \left\|(\psi^X)^{-1}\right\|$, where the inverse is the generalized
inverse (restricted to the support), for $X=B,C$. Then, for every $n$, 
\begin{align*}
%  \label{eq:H_min_lower}
  H_{\min}^\epsilon(A^n|B^n) &\geq n S(A|B) - (\mu_B+\mu_C)\sqrt{n\ln\frac{2}{\epsilon}}, \\
%  \label{eq:H_max_upper}
  H_{\max}^\epsilon(A^n|B^n) &\leq n S(A|B) + (\mu_B+\mu_C)\sqrt{n\ln\frac{2}{\epsilon}},
\end{align*}
and similar opposite bounds via Lemma~\ref{lemma:min-max-inequality}.
\altqed
\end{proposition}

\begin{lemma}[Proposition 5.5 in \cite{TomamichelThesis}]
\label{lemma:min-max-inequality}
Let $\rho \in \mathcal{S}(AB)$ and $\alpha,\beta \geq 0$ 
such that $\alpha+\beta < \frac{\pi}{2}$. Then,
\begin{equation*}
%\label{MinMaxIneq}
  H_{\min}^{\sin\alpha}(A|B)_{\rho} 
      \leq H_{\max}^{\sin\beta}(A|B)_{\rho} + \log \frac{1}{\cos^2(\alpha+\beta)}.
\end{equation*}
For $\epsilon,\delta \geq 0$, $\epsilon + \delta < 1$ this can be relaxed to the 
simpler form
\begin{equation*}
%\label{MinMaxIneq-simpler}
  \phantom{==:}
  H_{\min}^{\epsilon}(A|B)_{\rho} 
      \leq H_{\max}^{\delta}(A|B)_{\rho} + \log \frac{1}{1-(\epsilon+\delta)^2}.
      \phantom{==:}{\small\blacksquare}
\end{equation*}
%\altqed
\end{lemma}

\begin{lemma}[Dupuis~\cite{Fred:personal}]
\label{lemma:max-min-inequality}
Let $\rho \in \mathcal{S}(AB)$ and $0 \leq \epsilon \leq 1$. Then,
\begin{equation*}
%  \label{MaxMinIneq}
  H_{\max}^{\sqrt{1-\epsilon^4}}(A|B)_\rho \leq H_{\min}^{\epsilon}(A|B)_\rho,
\end{equation*}
which can be rewritten and relaxed into the form ($0 \leq \delta \leq 1$)
\begin{equation*}
  \phantom{=:}
%  \label{MaxMinIneq-simpler}
  H_{\max}^{\delta}(A|B)_\rho \leq H_{\min}^{\sqrt[4]{1-\delta^2}}(A|B)_\rho
                              \leq H_{\min}^{1-{\frac14}\delta^2}(A|B)_\rho.
  \phantom{=::}{\small\blacksquare}
\end{equation*}
\end{lemma}

%\begin{lemma}[Post-Selection Technique~\cite{CKR-post-select}]
%\label{lemma:post-select}
%For a Hilbert space $\cH$ of dimension $d$, denote by $\operatorname{Sym}^n(\cH)$
%the subspace of permutation-invariant states in $\cH^{\ox n}$. Then, for
%every state $\rho$ that is invariant under conjugation by 
%permutations, $\rho = \pi\rho\pi^\dagger$ for all $\pi\in S_n$, 
%\[
%  \rho \leq (n+1)^{d^2} \int {\rm d}\sigma\, \sigma^{\ox n},
%\]
%with a universal probability measure ${\rm d}\sigma$ on $\cS(\cH)$.
%\altqed
%\end{lemma}

%\begin{lemma}
%  \label{lemma:decouple-types}
%  ...
%\end{lemma}
%\begin{IEEEproof}
%Omitted.
%
% Well, of course in the full version we'll have to put it!
%
%\end{IEEEproof}

\section{Discussion}
\label{sec:discussion}
We showed that the min-entropic machinery employed in the
analysis of degradable quantum channels~\cite{DS05,CRS08,MorganWinter},
can be used equally, if not more easily, to obtain a \semistrong{}
converse for degraded quantum wiretap channels of the cqq kind;
the reason for focusing on this class of channels lies in the
availability of a single-letter formula. 

For degraded ccc-wiretap channels, i.e.~the original
Wyner model, Hayashi, Tyagi and Watanabe~\cite{HayashiTyagiWatanabe}
have found a very elegant argument, via hypothesis testing, to
give the tighter result that if $\epsilon+\delta < 1$, then
\(
  \lim_{n\rightarrow\infty} \frac1n \log M(n,\epsilon,\delta) = P(W).
\)
In fact, by phrasing error and privacy in terms of the trace distance 
$D(\rho,\sigma) = \frac12\|\rho-\sigma\|_1 \leq P(\rho,\sigma)$,
they show that this holds if and only if: For $1-\delta \leq \epsilon < 1$
the above limit gives the classical capacity $C(W)$ of the channel
from Alice to Bob.
It seems however that their
technique does not easily generalize to cqq-channels, as it 
exploits the classical nature of the output signals.

Similarly, if $\epsilon$ and $\delta$ are ``too big'', namely
$\sqrt{1-\delta^2} \leq \epsilon < 1$, we can easily
see that Theorem~\ref{thm:P-prettystrong} breaks: In that case,
\[
  \lim_{n\rightarrow\infty} \frac1n \log M(n,\epsilon,\delta) = C(W),
\]
where $C(W)$ is the classical capacity of the cq-channel
$W:x\longmapsto \rho_x^B$. Namely, choose any fixed $x_0^n$,
then for an asymptotically error-free and capacity-achieving 
code $x^n(m)$, with $m=1,\ldots, N=2^{nC(W)-o(n)}$, 
consider encoding message $m$ by the mixture
\(
  \epsilon^2 \proj{x_n(m)} + (1-\epsilon^2)\proj{x_0^n}.
\)
This scheme has transmission error arbitrarily close to 
$\epsilon$ and privacy error $\leq \sqrt{1-\epsilon^2} \leq \delta$.

By ignoring the privacy constraint, our Theorem~\ref{thm:P-prettystrong} 
includes a proof of the strong converse for the classical capacity of 
cq-channels~\cite{ON99,Winter99}; simply consider a trivial eavesdropper
and $\delta=0$. This shows that for $\epsilon < 1$, the 
limit of $\frac1n \log M(n,\epsilon,\delta)$ is bounded by $C(W)$;
cf.~Wang and Renner~\cite{WangRenner}.

\begin{figure}[ht]
\begin{center}
\includegraphics[width=7cm]{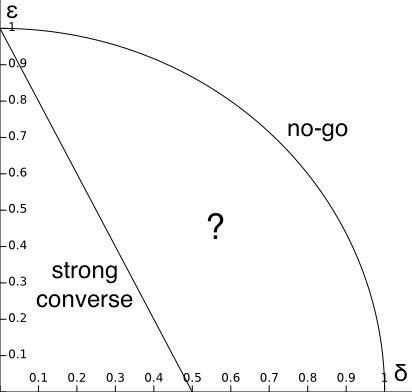}
\end{center}
\caption{Transmission error $\epsilon$ vs.~privacy error $\delta$: 
Below the straight line we have a strong converse, 
above the circle the strong converse cannot hold.
}
\end{figure}

We leave it as an open problem to try and close the gap between
the two regimes (see Fig.~1).

\section*{Acknowledgments}
It is a pleasure to thank Imre Csisz\'{a}r, Prakash Narayan,
Masahito Hayashi and Shun Watanabe for various discussions 
on wiretap channels, strong converses and the history of
cryptography.
The author was supported by the EU (STREP ``RAQUEL''), 
the ERC (Advanced Grant ``IRQUAT''), 
the Spanish MINECO (project project FIS2008-01236) with the support of FEDER funds, 
and the Generalitat de Catalunya (CIRIT project 2014-SGR-966).

%\vspace{0.5cm}

\end{document}